\documentclass[10pt,conference,letterpaper]{IEEEtran}
\usepackage{comment}
\usepackage{amsmath,amssymb}
\usepackage{booktabs}
\usepackage{cite}
\usepackage{graphicx}
\usepackage{xcolor}
\usepackage[hidelinks]{hyperref}
\usepackage{algorithm}
\usepackage{algpseudocode}
\graphicspath{{figures/}}
\usepackage{multirow}
\title{StreamAhead: Real-Time Streaming Video \\ Question Answering via Adaptive Precomputation}
\author{Haoyang Fan$^{1}$ \quad
Dhruv Parikh$^{1}$ \quad
Rajgopal Kannan$^{2}$ \quad
Viktor Prasanna$^{1}$\\
$^{1}$University of Southern California (USC), Los Angeles, CA, USA\\
$^{2}$DEVCOM Army Research Office\\
{\tt\small \{haoyangf,dhruvash,prasanna\}@usc.edu}\\
{\tt\small rajgopal.kannan.civ@army.mil}
}

\begin{document}

\maketitle

\begin{abstract}

Real-time streaming video question answering with vision--language models must
answer queries that arrive at any time with low latency while sustaining the
video-processing rate required by the application. Existing systems construct
key--value (KV) caches for the entire video as it arrives, although a query
typically uses only a few of the preceding frames as evidence. This highlights a fundamental trade-off: performing more
computation as video arrives reduces query-time computation but can increase
processing lag, whereas deferring computation can increase the video-processing
rate but requires additional query-time computation for retrieved content.

We present StreamAhead, an algorithm--system co-design that exploits this
trade-off through adaptive precomputation: dynamically controlling how much
computation to complete for each video segment before a query arrives.
StreamAhead encodes video segments independently and lets each remain at a
different representation level, from inexpensive visual features to costly KV
caches, completing KV caches only for the retrieved segments when a query
arrives. 
StreamAhead profiles offline the cost of advancing a segment to each
representation level. Its runtime uses these profiles to schedule segment
computation concurrently with query execution, pipelining it across
heterogeneous CPU and GPU resources while maintaining the required
video-processing rate.
StreamAhead achieves accuracy competitive with state-of-the-art methods with up
to 10.4$\times$ lower time-to-first-token (TTFT), and up to 5.9$\times$ lower
TTFT than state-of-the-art streaming systems. It sustains 30\,FPS with P95 TTFT
under 0.5\,s, and up to 57\,FPS with P95 TTFT under 2\,s. To our knowledge,
StreamAhead is the first real-time streaming video QA system with competitive
accuracy, setting a new Pareto frontier between video-processing rate and TTFT.
\end{abstract}



\section{Introduction}
\label{sec:introduction}

\begin{figure*}[t]
    \centering
    \includegraphics[width=\textwidth]{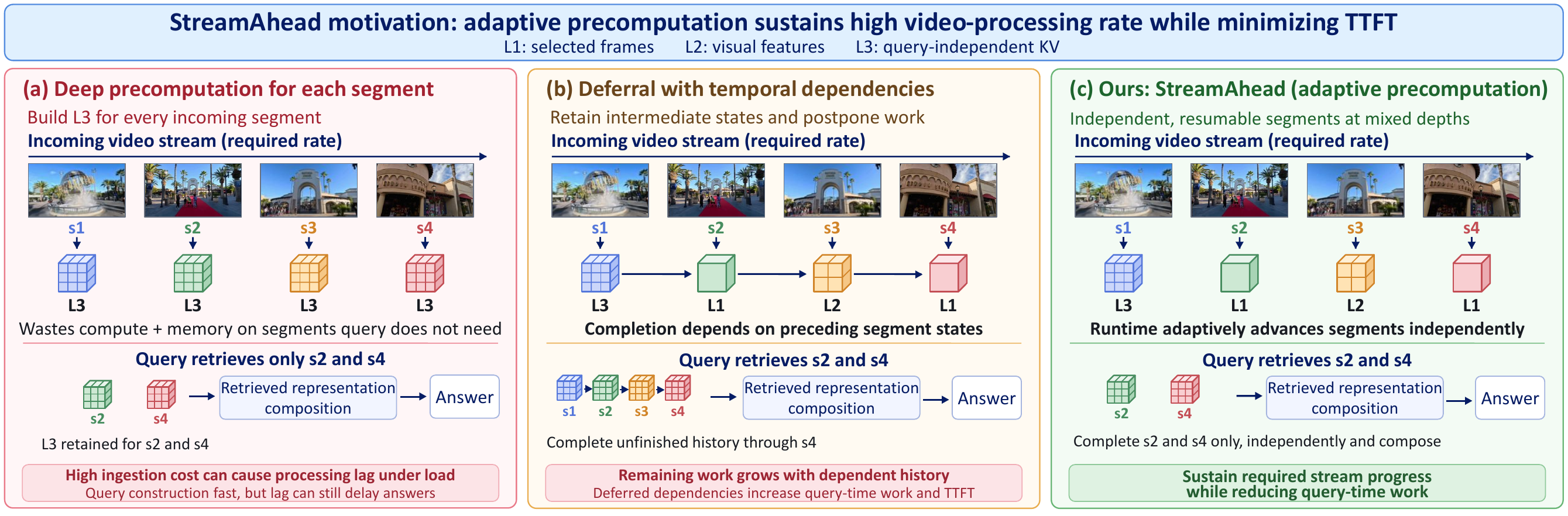}
    \caption{(a) Building KV caches for every segment, (b) deferring computation under temporal dependencies, and (c) adaptive precomputation in StreamAhead.}
    \label{fig:motivation}
\end{figure*}

Streaming video question answering with vision--language models (VLMs)
considers a continuously arriving video stream in which users may issue
questions at arbitrary times~\cite{videollmonline2024,rekv2025,ovobench,streamingbench,egolife2025,egoextra2025}.
An application specifies the video sampling rate and resolution that the system
is expected to process. The system must sustain this processing rate;
otherwise, a growing backlog creates \emph{processing lag}, i.e., a delay
between video arrival and completion of the required processing~\cite{videostorm2017}, making recent
video unavailable when a query needs it. In this paper, we therefore study a
fundamental question for streaming video question answering: \emph{how can we
sustain the required video-processing rate while simultaneously minimizing
time-to-first-token (TTFT), measured from query arrival to the first generated
token?} We call a system real-time when it achieves \emph{both}: it keeps pace with
the incoming video and answers queries with low latency.

Existing streaming video QA systems reduce query-time computation by
precomputing reusable video representations, such as visual features or
key--value (KV) caches, that can later be reused when relevant historical
content is retrieved~\cite{flashvstream2025,rekv2025,streamkv2025,livevlm2026,mukv2026,hermes2026}. For example, ReKV~\cite{rekv2025} incrementally constructs reusable KV states for incoming
video and retrieves query-relevant cached states for answer generation.
However, constructing deeper representations such as video KV caches requires
substantial computation before the relevance of that content to a future
query is known. A particular query typically depends on only a small subset of
the retained video history, which can be identified through retrieval
mechanisms~\cite{rekv2025,streamkv2025,goldfish2024,videoragluo2025}. Fully constructing deep
representations for the remaining history therefore spends processing capacity
on content that may never be used by that query. This extra computation can cause the system to fall behind the incoming stream and accumulate processing lag, without benefiting the query.

Rather than constructing full KV caches for the entire stream as video
arrives, a system can retain intermediate results and postpone the
remaining computation~\cite{focus2018,boggart2023,shallowstream2026}. This exposes a fundamental trade-off: performing more
computation as video arrives reduces query-time computation but can increase
processing lag, whereas deferring computation increases the video-processing
rate but requires additional computation at query time if that content is
later retrieved.
This motivates what we call \emph{adaptive precomputation}: deciding at
runtime how much computation to complete before a query arrives, based
on the current workload and available resources.

Exploiting this trade-off, however, raises two challenges. \emph{First}, the
system must be able to complete different amounts of computation for different
parts of the video. Many existing systems build the KV caches of new frames
from the cached states of earlier
frames~\cite{rekv2025,streamkv2025,streamingvlm2026,videostreaming2024,livevlm2026}.
Under this temporal dependency, completing one retrieved segment at query time
first requires completing every earlier segment. The query-time cost then grows
with the length of the video history, not with the retrieved content required as evidence for that query.
\emph{Second}, the system must decide at runtime which segments to precompute
and how much computation to complete for each. It must also decide when to run
this work, since video processing, query answering, and precomputation share
the same compute. Finally, it must place the results across CPU and GPU memory, as GPU
memory is limited and loading results back from CPU memory delays the query.
Throughout, the system must not fall behind the incoming video stream.

To address these challenges, we present StreamAhead, an
algorithm--system co-design for adaptive precomputation in streaming
video QA (Figure~\ref{fig:motivation}). StreamAhead divides the incoming video into fixed-duration
segments and precomputes each segment independently, without any temporal
dependency on other segments' KV caches. Each segment retains results at one of three
\emph{representation levels}, corresponding to successive stages of
processing that segment: raw frames and a retrieval key (L1), visual features
(L2), or a query-independent KV cache (L3). Later processing continues from the
stored results without repeating completed stages. Before a query arrives, different segments can
remain at different levels according to the available resources.
At query time, StreamAhead retrieves relevant segments, reuses
existing KV caches, and completes KV construction for retrieved
segments still at L1 or L2. StreamAhead then combines the retrieved KV caches
through \emph{retrieved representation composition}, placing them in video
order with continuous positions to form a single
context~\cite{blockattention2025,ape2025}. Because segments never attend to one
another, we also propose a lightweight mechanism to recompute each segment's
short text suffix in this context, adding cross-segment interaction without
rebuilding visual KV states.

To address the second challenge, StreamAhead first profiles offline the
compute and memory cost of advancing a segment to each representation level.
It also profiles query TTFT over segments at each level, and uses both profiles
to select the processing configuration and required representation level for
the target workload~\cite{videostorm2017,chameleon2018}.
Online, a shared segment table tracks each segment's completed representation
level and whether its results are in CPU or GPU memory. It also tracks ongoing
work on the segment, such as advancing it to a deeper level or moving it
between CPU and GPU. With this table, video processing and query execution
run concurrently: a query that needs a segment already being advanced reuses
that work instead of repeating it. The runtime pipelines segment
computation across heterogeneous CPU and GPU resources, overlapping CPU
preparation and data movement with GPU execution. It also places
representations across CPU and GPU
memory~\cite{infinigen2024,cachedattention2024,hcache2025}. Its scheduler keeps
pace with the incoming video, gives priority to query work when a query
arrives, and uses spare capacity for additional precomputation. While a query
runs, processing of newly arriving video may fall behind. After the query, the
runtime first clears this backlog and only then resumes additional
precomputation. We make the following contributions:
\begin{itemize}
    \item \textbf{Algorithm.} We process video as independent segments with
    three reusable representation levels, so that different amounts of
    computation can be completed for each segment before a query arrives. We
    combine retrieved segments through retrieved representation composition,
    with a lightweight mechanism that adds cross-segment interaction.

    \item \textbf{Scheduling.} We design profile-guided scheduling for adaptive
    precomputation. Offline profiling selects the required representation
    level. Online scheduling keeps pace with the incoming video, prioritizes
    queries, and uses spare capacity for additional precomputation.

    \item \textbf{Runtime.} We build a runtime that executes video processing
    and queries concurrently. It uses a shared segment table, pipelining
    across heterogeneous CPU and GPU resources, and a CPU--GPU memory
    hierarchy.
\end{itemize}

StreamAhead achieves accuracy competitive with state-of-the-art methods with up
to 10.4$\times$ lower TTFT, and up to 5.9$\times$ lower TTFT than
state-of-the-art streaming systems. It sustains 30\,FPS with P95 TTFT under
0.5\,s, and up to 57\,FPS with P95 TTFT under 2\,s. To our knowledge,
StreamAhead is the first real-time streaming video QA system with competitive
accuracy, setting a new Pareto frontier between video-processing rate and
TTFT.

\section{Related Work}
\label{sec:related}

\subsection{Streaming Video Question Answering}
\label{sec:rw-streaming-qa}

Streaming video question answering requires a model to retain information from a continuously
arriving video and answer questions issued at arbitrary times~\cite{ovobench,streamingbench}.
Early work trains online video assistants that learn when to respond and what to remember as
frames arrive~\cite{videollmonline2024,videostreaming2024,dispider2025,streambridge2025}.
Memory-based designs summarize the video history into compact visual features, events, or
text~\cite{flashvstream2025,streamchat2025,oasis2026,vista2026}, while retrieval-based approaches
select the parts of a long video that are relevant to a
question~\cite{goldfish2024,videoragluo2025,aks2025}. Closest to our work, several methods retain
the video history as KV caches, so that language-model computation can be reused when relevant
content is retrieved~\cite{rekv2025,streamkv2025,livevlm2026,mukv2026}, or maintain a bounded KV
cache within a fixed memory budget~\cite{streamingvlm2026,infinipotv2025,hermes2026,dscache2026}.

These works show that precomputing reusable video representations substantially reduces
query-time computation. In most of them, however, the KV caches of new frames are constructed from
the cached states of earlier frames. The computation completed as video arrives is also fixed by
design: every incoming frame is encoded into the KV cache whether or not a later query uses it. Moreover, many of these
methods are evaluated on answer quality with the video processed offline, without requiring that
the incoming video be processed as fast as it arrives. Under real-time streaming, this fixed
computation can fall behind the incoming video and accumulate processing lag, making recent video
unavailable when a query needs it. StreamAhead instead processes video segments independently and
decides at runtime how much computation to complete for each segment before a query arrives,
i.e., adaptive precomputation. It thereby keeps pace with the incoming video while completing KV
caches only for the retrieved segments at query time.

\subsection{Efficient Streaming Video Inference}
\label{sec:rw-efficient}

A complementary line of work reduces the cost of processing each frame so that a system can keep
pace with the incoming video. Token pruning and compression methods reduce the number of visual
tokens processed by the language model~\cite{fastv2024,visionzip2025,longvu2025,videoxl2025}, and
streaming variants apply similar reductions as frames arrive~\cite{streamingtom2026,stc2026}.
Other work invokes the language model only when relevant events occur~\cite{streammind2025},
accelerates KV retrieval during streaming prefill with dedicated hardware~\cite{vrex2026}, or
reuses computation across overlapping video windows~\cite{codecsight2026}. Recent systems also
treat video-processing throughput and query latency as explicit goals: one pipelines model stages
to sustain high frame rates, evaluating TTFT separately~\cite{vicostream2026}, and another precomputes only the
first layers of the language model as video arrives and completes the remaining layers at query
time~\cite{shallowstream2026}. Evaluation protocols have likewise begun to measure latency under
realistic video arrival~\cite{streamingeval2026}.

These advances make each unit of computation cheaper or better overlapped. They either reduce the
visual evidence processed by the model or fix how much computation is completed before a query
arrives, such as a full KV cache for every video chunk or a fixed number of layers. StreamAhead
keeps the visual evidence unchanged and adapts the amount of precomputation to the
current workload: when spare capacity is available, segments advance to KV caches before a query
arrives, and under heavier load they remain at earlier representation levels so that processing
lag does not accumulate. Token reduction is orthogonal to this decision and can be applied within each
segment.

\subsection{Scheduling Under Rate and Latency Requirements}
\label{sec:rw-scheduling}

Systems for live video analytics have long scheduled computation to keep pace with incoming
streams. They adapt processing configurations to bound processing lag under limited
resources~\cite{videostorm2017,chameleon2018}, schedule inference across GPUs to meet latency
targets~\cite{nexus2019}, and share accelerators between work that must keep pace with the stream
and background work~\cite{ekya2022}. Video query systems split computation between inexpensive
processing as video arrives and expensive processing when a query is
issued~\cite{focus2018,boggart2023}. In LLM serving, systems meet latency targets through
iteration-level batching, paged KV memory, chunked prefill, and disaggregated prefill and
decoding~\cite{orca2022,vllm2023,sarathiserve2024,distserve2024,splitwise2024}, and multimodal
serving systems separate or overlap visual encoding and language-model
prefill~\cite{modserve2025,epd2025,rserve2025}. Related techniques keep KV caches across CPU and
GPU memory~\cite{infinigen2024,cachedattention2024,hcache2025}, use idle GPU capacity for
deferrable work~\cite{conserve2026}, or overlap the arrival of streaming input with
prefill~\cite{stream2llm2026}.

StreamAhead builds on these ideas with two differences. Unlike video analytics systems, its
precomputation produces reusable model states, and how much work is completed as video arrives
versus at query time is decided at runtime for each segment rather than fixed in advance. Unlike
LLM serving systems, its input is a continuously arriving video that must be processed at the
required rate while queries are answered concurrently. StreamAhead's scheduler therefore gives priority to
keeping pace with the incoming video and answering queries, and uses spare capacity to precompute
KV caches for segments that future queries may retrieve.

\section{Problem Definition}
\label{sec:problem}

\noindent\textbf{Streaming input.} We consider the incoming video as a sequence of fixed-duration
segments $s_1, s_2, \ldots$ of duration $\Delta$, where segment $s_i$ becomes available at time
$t_i = i\Delta$. The application specifies the video sampling rate and resolution. Each segment
therefore contains $K$ frames to be processed, and the required video-processing rate is
$r = K/\Delta$ frames per second.

\noindent\textbf{Processing lag.} As segments arrive, the system performs their \emph{required
processing}, which takes time $T_{\mathrm{req}}$ per segment. The system keeps pace with the
incoming video if $T_{\mathrm{req}} \le \Delta$. Otherwise, each new segment adds
$T_{\mathrm{req}} - \Delta$ to the processing lag, i.e., the delay between a segment's arrival and
the completion of its required processing. The lag then keeps accumulating as the stream continues.

\noindent\textbf{Query latency.} A query $q$ is issued at time $t_q$, unknown in advance. A VLM
answers it from the segments that have arrived by $t_q$, and the answer may depend on any of them.
Its time-to-first-token, $\mathrm{TTFT}(q)$, is measured from $t_q$ to the first generated token.
It includes the computation still needed for the segments the query uses. Processing lag increases TTFT, because segments whose processing has
fallen behind need more computation when a query uses them. As lag accumulates, more of the
arrived video is affected. Existing streaming video QA methods typically measure TTFT on video that
has already been processed, and therefore do not account for processing lag.

\noindent\textbf{Real-time streaming video QA.} We call a system \emph{real-time} at rate $r$ with
TTFT target $\tau$ if it keeps pace with the incoming video, $T_{\mathrm{req}} \le \Delta$, and
answers every query with $\mathrm{TTFT}(q) \le \tau$. The goal is to remain real-time at the highest
video-processing rate, without reducing the sampling rate or resolution and without knowledge of
future queries. The two conditions compete for the same resources: processing performed as
segments arrive adds to $T_{\mathrm{req}}$, whereas processing deferred until a query arrives adds
to its TTFT.

\section{Method}
\label{sec:method}

\begin{figure*}[t]
    \centering
    \includegraphics[width=0.9\textwidth]{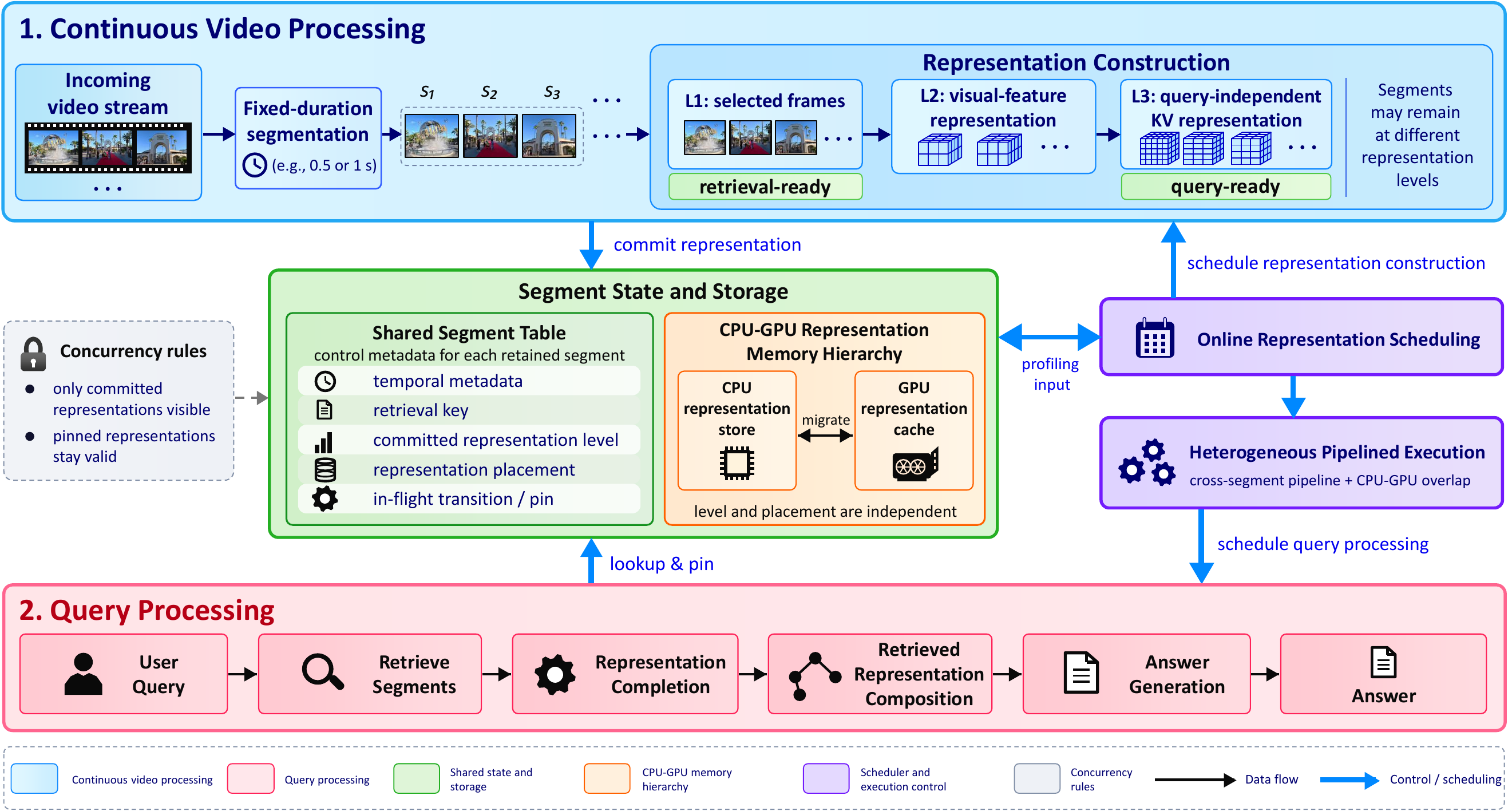}
    \caption{Overview of StreamAhead: continuous video processing, shared segment state and storage, and query processing.}
    \label{fig:system_overview}
\end{figure*}

As defined in Section~\ref{sec:problem}, StreamAhead balances $T_{\mathrm{req}}$
against TTFT through adaptive precomputation over independent segments at
multiple representation levels.
Before a query arrives, different segments may complete different amounts of
computation. When a query is issued, StreamAhead retrieves
a small set of relevant segments, reuses the deepest committed representation
available for each segment, completes only the remaining computation required
by those segments, and composes the resulting representations for answer
generation. We use \emph{precomputation} to denote the computation completed
for each segment before a query arrives. Figure~\ref{fig:system_overview} gives an overview of StreamAhead.

\subsection{Independent Segment Processing}
\label{sec:independent-segments}

StreamAhead partitions the continuous video stream into fixed-duration
segments $s_1,s_2,\ldots,s_n$. Each segment is processed under the same fixed
textual prefix, but its visual representation and query-independent
language-model state are constructed without using state from
other segments. This independence is essential for adaptive
precomputation: if a segment depends on reusable states constructed from
earlier video, deferring its computation may also require completing the
preceding dependent history before the segment can resume computation. StreamAhead
removes this cross-segment dependency, allowing each segment to advance,
remain at an intermediate representation, or resume later without requiring
neighboring segments to reach the same processing stage. Computed
segment states therefore serve as the basic units of precomputation
and runtime scheduling.

Independent construction is not equivalent to jointly encoding multiple
segments because the reusable state of one segment has not attended to the
visual content of other segments. StreamAhead accounts for this distinction
when retrieved segment representations are combined at query time.

\subsection{Representation Levels and Precomputation}
\label{sec:representation-levels}

For each segment $s_i$, StreamAhead maintains one committed reusable
representation at one of three levels:
\[
\ell_i \in \{L1,L2,L3\},
\qquad
L1(s_i)
\xrightarrow{C_{1\rightarrow2}}
L2(s_i)
\xrightarrow{C_{2\rightarrow3}}
L3(s_i),
\]
where $C_{1\rightarrow2}$ and $C_{2\rightarrow3}$ denote the computation
required by the corresponding representation transitions. $L1$ contains the
video frames and their temporal metadata, $L2$ contains the visual
features and associated model inputs required to continue language-model
processing without repeating visual encoding, and $L3$ contains the
query-independent KV representation together with the token and position
metadata required for later query-time composition. All three levels
correspond to the same visual evidence: advancing to a deeper level
completes additional query-independent computation.

A retrieval key $k_i$ is constructed when $L1$ becomes available and is
retained separately. Once $L1$ and $k_i$ are committed, the
segment becomes available for retrieval, and the key remains unchanged as the
reusable representation advances through $L2$ and $L3$. The committed
representation level therefore records how much computation has
been completed and determines how much computation would remain if
the segment were later retrieved. Different segments may remain at different
representation levels before query arrival; the runtime policy that determines
when these transitions execute is described in
Section~\ref{sec:scheduling}. Precomputation changes only how much
computation is completed for each segment. It never changes the total frames the
model sees or their resolution.

\begin{algorithm}[t]
\caption{StreamAhead Segment and Query Processing}
\label{alg:streamahead}
\begin{algorithmic}[1]

\Procedure{IndexSegment}{$s_i$}
    \State Construct $L1(s_i)$ and retrieval key $k_i$
    \State Commit $L1(s_i)$, $k_i$, and temporal metadata
\EndProcedure

\Procedure{AdvanceSegment}{$s_i$}
    \If{$\ell_i=L1$}
        \State Construct and commit $L2(s_i)$ from $L1(s_i)$
    \ElsIf{$\ell_i=L2$}
        \State Construct and commit $L3(s_i)$ from $L2(s_i)$
    \EndIf
\EndProcedure

\Procedure{AnswerQuery}{$q,t_q$}
    \State $\mathcal{R}_q \gets \Call{Retrieve}{q,t_q}$
    \State Order $\mathcal{R}_q$ by video time
    \ForAll{$s_i \in \mathcal{R}_q$}
        \State Reuse the deepest committed representation of $s_i$
        \State Complete $s_i$ to $L3$ if necessary
    \EndFor
    \State Remap positions and compose retrieved $L3$ representations
    \State Append $q$ and generate the answer
\EndProcedure

\end{algorithmic}
\end{algorithm}

\subsection{Query-Time Completion and Retrieved Representation Composition}
\label{sec:query-processing}

For a query $q$ issued at time $t_q$, let $\mathcal{I}_q$ denote the
segments whose retrieval keys have been committed by $t_q$, and let
$\mathcal{R}_q \subseteq \mathcal{I}_q$ denote the segments selected by
retrieval. Retrieved segments are ordered by their
original video timestamps. For a retrieved segment $s_i$, the remaining
representation construction is
\[
C_{\mathrm{remain}}(s_i)=
\begin{cases}
C_{1\rightarrow2}(s_i)+C_{2\rightarrow3}(s_i), & \ell_i=L1,\\
C_{2\rightarrow3}(s_i), & \ell_i=L2,\\
0, & \ell_i=L3.
\end{cases}
\]
StreamAhead therefore reuses the deepest committed representation available
for each retrieved segment and performs only $C_{\mathrm{remain}}(s_i)$ on
the query critical path. Precomputation performed for segments outside
$\mathcal{R}_q$ does not enter the current query's critical path.

Once the retrieved segments are available at $L3$, StreamAhead constructs a
temporary context from their independently computed representations.
The segments are placed in temporal order, their segment-local token positions
are remapped to the composed sequence, and the cached keys are adjusted
according to the resulting positional encoding. StreamAhead then combines the retrieved states and can recompute each
segment's short text suffix to add cross-segment interaction.
The composed context does not modify the persistent representations retained
for future queries. Algorithm~\ref{alg:streamahead} summarizes segment and query processing. Retrieval and composition details are described in Section~\ref{sec:implementation}.

\section{System Design}
\label{sec:design}

\begin{figure*}[t]
\centering
\includegraphics[width=\textwidth]{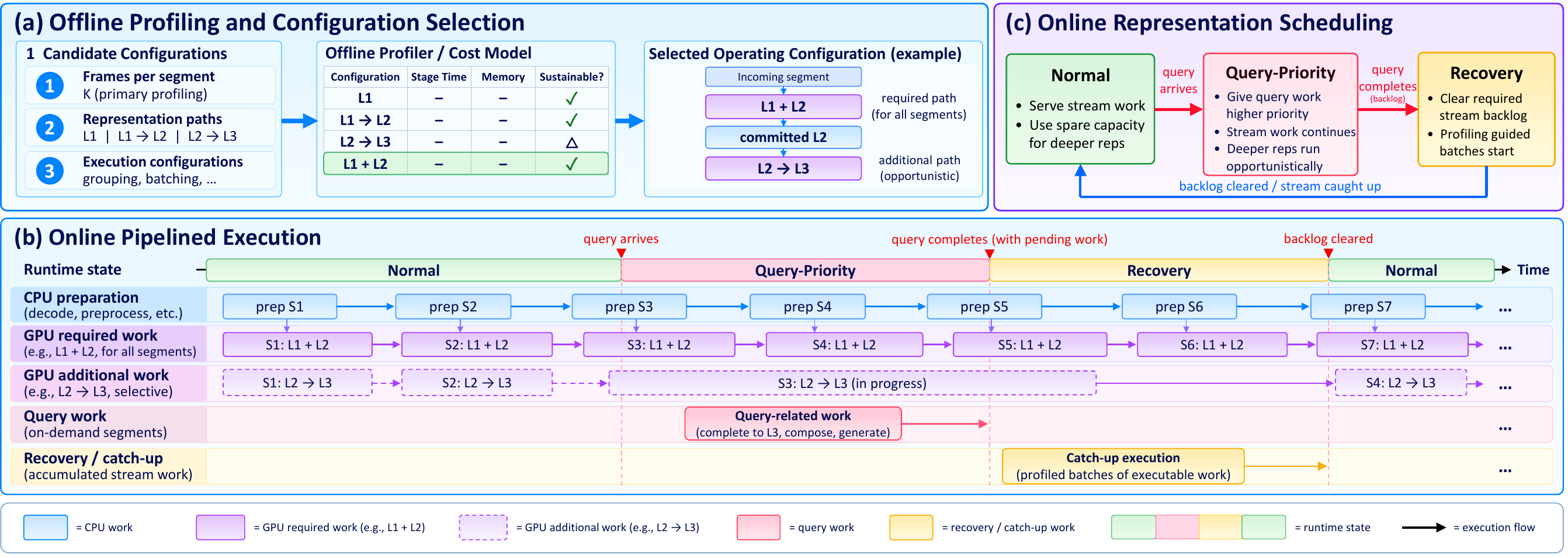}
\caption{Offline profiling, online pipelined execution, and runtime scheduling states in StreamAhead.}
\label{fig:timeline}
\end{figure*}

StreamAhead builds on independent segment processing and representation levels to support adaptive precomputation. A shared segment table maintains each segment's committed representation, placement, and in-flight state, providing a consistent view for concurrent video processing and query execution. The system combines offline profiling with an online runtime (Figure~\ref{fig:timeline}). Offline profiling characterizes computation, query-time, and memory costs to select a processing configuration and required representation level. During online execution, the scheduler uses these profiles and the shared segment state to select representation transitions and assign execution priorities. A heterogeneous CPU--GPU pipeline executes the scheduled work through cross-segment pipelining and CPU--GPU overlap, while reporting execution progress through the shared table. Further, a CPU--GPU representation memory hierarchy manages the placement of completed representations under bounded memory.

\subsection{Offline Profiling and Configuration Selection}
\label{sec:offline-profiling-design}

StreamAhead profiles continuous video processing and query-time execution before online execution. For a fixed model, resolution, segment duration, retrieval configuration, and composition method, the profiler varies the number of frames processed per segment, $K$ (Section~\ref{sec:problem}). For each candidate $K$, the profiler measures the wall-clock latency of L1 and retrieval-key construction, L1-to-L2 visual encoding, and L2-to-L3 KV construction. It also records the corresponding L2 and L3 memory footprints. Using representations constructed under the same configuration, StreamAhead measures end-to-end TTFT over segments at each level, and the major query-time costs associated with representation movement and composition. These profiles characterize how the visual workload affects both continuous processing and subsequent query execution.
Configuration selection uses these measurements together with the required video-processing rate, TTFT target, available memory, and answer-quality measurements to select the processing configuration. The selected configuration fixes the visual workload and required representation level for incoming segments. The required level must be maintained for every incoming segment, while advancement beyond this level remains an online scheduling decision.

\subsection{Shared Segment Table}
\label{sec:shared-state}

StreamAhead maintains a shared record for every retained video segment. Each record contains the segment's temporal metadata and retrieval key, its committed representation level (the latest completed level) and representation placement (CPU or GPU memory), any in-flight transition (work currently advancing or moving the segment), and whether the committed representation is currently in use. Video processing, precomputation, query execution, and memory management consult the same record when operating on a segment. Only committed representations are visible to other tasks. When a representation transition begins, the previously committed representation remains available until the new representation has been fully constructed. For example, a segment undergoing an L2-to-L3 transition remains committed at L2 until the transition completes, after which the new L3 representation becomes visible. Model execution and data movement occur outside the table update, so no task sees a partially constructed representation. The table also records in-flight transitions so that concurrent tasks do not duplicate the same representation construction. If a query requires a transition that is already in progress, it can reuse or wait for that work. Representations used by an active query are pinned until the query releases them, preventing migration or reclamation during use. Retrieval keys remain unchanged as representation level and placement change, keeping retrieval independent of runtime representation management.

\subsection{Online Representation Scheduling}
\label{sec:scheduling}

After the processing configuration is fixed, StreamAhead schedules representation transitions to keep pace with the incoming video while using available capacity for additional precomputation. For the selected configuration, transitions needed to reach the required representation level constitute required video processing, while transitions beyond this level are additional precomputation. The scheduler uses offline execution profiles and the shared segment table to select transitions and assign execution priorities. Required video processing takes precedence over additional precomputation, limiting interference with the processing of newly arriving segments. StreamAhead operates in three runtime states: \emph{Normal}, \emph{Query-Priority}, and \emph{Recovery}.

\subsubsection{Normal}
Newly arriving segments are prioritized to keep pace with the incoming video and prevent processing lag from accumulating. When the required work is caught up, available execution capacity is used to advance segments toward deeper representations. The scheduler selects transitions according to their committed and in-flight states, avoiding duplicate computation.

\subsubsection{Query-Priority}
When a query arrives, computation required to complete the retrieved segments and generate the answer becomes latency-critical and runs first. Newly arriving video continues to be processed, while additional precomputation receives lower priority. Already executing transitions are not explicitly aborted or restarted, and compatible in-flight computation remains reusable. Query execution may temporarily reduce the resources available for incoming video processing.

\subsubsection{Recovery}
If required video processing accumulates during query execution, StreamAhead enters Recovery to clear this backlog. Rather than simply resuming normal execution, the scheduler uses the current backlog and offline profiles to select an efficient batch size. When multiple compatible transitions are already ready, it may select batched execution to improve backlog processing throughput under the available resource constraints. No additional waiting is introduced solely to form a batch. Once the required backlog is cleared, StreamAhead returns to Normal execution and resumes additional precomputation.

\subsection{Heterogeneous Pipelined Execution}
\label{sec:pipelined-execution}

StreamAhead executes scheduled representation transitions through a heterogeneous CPU--GPU pipeline. The scheduler selects the work and its execution priority, while the pipeline provides concurrent execution across segments and processing stages. Execution progress is recorded in the shared segment table and made available for subsequent scheduling decisions.

\subsubsection{Cross-segment pipelining}
Segment independence allows different segments to occupy different stages, such as L1$\rightarrow$L2 and L2$\rightarrow$L3, concurrently. A segment does not need to reach the deepest representation level before subsequent segments begin processing. This enables incoming video processing and additional precomputation to overlap while preserving the required computation order within each segment.

\subsubsection{CPU--GPU overlap}
StreamAhead pipelines CPU-side frame preparation, metadata processing, and data movement with GPU model execution across segments. Preparation and data movement for one segment can overlap with GPU computation for another, reducing GPU idle time.

\subsubsection{Concurrent execution support}
The execution pipeline supports concurrent GPU work at different priorities and batched execution of compatible representation transitions. The scheduler assigns priorities and selects batch sizes from the execution profiles, while the pipeline executes the scheduled work and reports completion through the shared segment table.

\subsection{CPU--GPU Representation Memory Hierarchy}
\label{sec:memory-hierarchy}

StreamAhead manages representation level and representation placement independently. For segment $s_i$, let $\ell_i \in \{L1,L2,L3\}$ and $p_i \in \{\mathrm{CPU},\mathrm{GPU}\}$ denote its committed representation level and current placement, respectively. The representation level records completed computation, whereas placement records where the committed representation is stored. StreamAhead uses a CPU representation store for retained representations and a bounded GPU representation cache for representations needed on the GPU. Migration changes placement without changing the committed level. For example, $(L3,\mathrm{GPU}) \rightarrow (L3,\mathrm{CPU})$ preserves the completed L3 representation, which can later be loaded back to the GPU without repeating visual encoding or KV construction. Each segment keeps one committed representation. A transition or migration may temporarily require both the current representation and a destination allocation until the operation completes. Representations pinned by active queries cannot be migrated or reclaimed until those queries release them. Before admitting new work, StreamAhead accounts for committed representations, transition outputs, data movement, and temporary execution workspace. When the GPU representation cache reaches its capacity, unpinned representations, least recently used first, can migrate to the CPU representation store while retaining their committed levels. Query-time position remapping, representation composition, and answer generation operate on temporary per-query state and do not modify the persistent representation stored for a segment.

\section{Implementation}
\label{sec:implementation}

This section presents the implementation of StreamAhead's algorithmic components and CPU--GPU runtime.

\subsection{Algorithmic Realization}
\label{sec:impl-algorithm}

For each fixed-duration segment, $L1$ retains the frames and temporal metadata, $L2$ stores visual-encoder outputs and the inputs to the language model, and $L3$ stores query-independent KV states and position metadata. A SigLIP image encoder~\cite{siglip2023} constructs the retrieval key alongside $L1$. At query time, the SigLIP text encoder produces a question embedding, and retrieval combines the newest segment with semantically ranked historical segments using cosine similarity. For representation composition, StreamAhead retains one shared textual prefix, orders retrieved segments by video time, remaps their positions, and applies MRoPE relocation~\cite{qwen3vl2025} to cached keys. \emph{Concat} directly assembles the positioned KV caches, while \emph{TailRepair}, our lightweight mechanism, recomputes each segment's short text suffix in the composed context. Both operate on temporary per-query state without modifying persistent representations.

\subsection{Shared Segment State}
\label{sec:impl-state}

Each retained segment has a CPU-resident record in the shared segment table containing its temporal metadata, retrieval key, committed representation level, placement, storage handle, in-flight transition, and active-use state. Transitions follow a begin--compute--commit sequence: computation and data movement occur outside the table update, and the new representation becomes visible only after completion. Queries reuse or wait for compatible in-flight transitions and pin the representations they use to prevent migration or reclamation during use.

\subsection{Representation Storage and Query Access}
\label{sec:impl-memory}

StreamAhead uses a bounded GPU representation cache for recently used $L3$ representations and a larger CPU representation store for $L1$ data, persistent $L2$ representations, and evicted $L3$ representations.
Migration preserves the committed representation level. Before a transition or migration, the runtime reserves memory for the required allocations and execution workspace, and previous storage is reclaimed only after the new representation is committed and no query still uses it. At query time, GPU-resident $L3$ is reused directly, CPU-resident $L3$ is transferred to the GPU, and shallower representations are completed to $L3$ as needed. Temporary per-query state is released after generation, and the associated representations are unpinned.

\subsection{Concurrent Execution Runtime}
\label{sec:impl-runtime}

StreamAhead uses asynchronous CPU workers and CUDA streams to implement cross-segment pipelining and CPU--GPU overlap. GPU work is assigned according to its runtime role rather than a fixed representation level: required video processing and query-critical work use the higher-priority execution path, while additional precomputation uses the lower-priority path. CUDA stream priority provides preferential scheduling at kernel boundaries without explicitly preempting in-flight representation transitions. During Recovery, compatible ready transitions may use profiled batched execution when beneficial, without waiting for future segments to form a batch. The offline profiler measures costs using the same execution paths to support configuration selection and recovery.


\section{Evaluation}
\label{sec:evaluation}

\subsection{Experimental Setup}
\label{sec:experimental-setup}

\noindent\textbf{Benchmarks and models.}
We evaluate answer quality and algorithmic sensitivity
on OVO-Bench (Perception and Backward) and StreamingBench
(Real-Time). Videos are temporally sampled at the evaluated video-processing rate, with source frames repeated when the requested rate exceeds the source frame rate.
We evaluate Qwen2.5-VL-7B and Qwen3-VL-8B,
comparing answer quality with published results.
For system comparisons, we select three representative
streaming video QA methods: ReKV for KV-cache retrieval
and reuse, StreamingVLM for incremental streaming
inference with temporally dependent KV states, and
OASIS for algorithmic optimization through hierarchical
video memory and retrieval.
These methods represent complementary approaches to
streaming video processing and query serving.

\noindent\textbf{System configuration.} Unless specified otherwise, system experiments use Qwen3-VL-8B with BF16, FlashAttention-2, and $448$-pixel visual inputs. Each segment maintains a separate retrieval key, and the default retrieval policy selects the latest segment plus two semantically ranked historical segments. Offline profiling selects the required representation level for each rate. In the rate sweep, the required level is L3 below 30\,FPS, L2 from 30 to 80\,FPS, and L1 above 80\,FPS. Beyond the required level, the runtime performs additional L3 precomputation when capacity allows. The online replay comparison uses L2 as the required level. We use a 16-GiB GPU representation cache. 

\noindent\textbf{Hardware and scope.} Experiments run on a server with two AMD EPYC 9654 CPUs, 1.5\,TiB of host memory, and four NVIDIA RTX 6000 Ada GPUs (48\,GiB each). All experiments use just one GPU.

\subsection{Answer Quality and Sensitivity}
\label{sec:answer-quality}

\begin{table}[t]
\centering
\caption{Answer quality on OVO-Bench (accuracy, \%).}
\label{tab:accuracy-ovo}
\footnotesize
\setlength{\tabcolsep}{4pt}
\renewcommand{\arraystretch}{1.0}
\resizebox{0.9\columnwidth}{!}{%
\begin{tabular}{lrr}
\toprule
Method & Perception & Backward \\
\midrule
Gemini 1.5 Pro & 69.3 & 62.5 \\
GPT-4o & 64.5 & 60.8 \\
LLaVA-Video-7B & 62.3 & 53.7 \\
LLaVA-OneVision-7B & 62.0 & 43.7 \\
Flash-VStream-7B & 28.4 & 27.4 \\
VideoLLM-online-8B & 22.2 & 16.2 \\
StreamChat & 50.1 & 47.4 \\
Dispider & 54.6 & 36.1 \\
ReKV (LLaVA-OV-7B) & 64.4 & 64.6 \\
StreamingVLM (Qwen2.5-VL-7B) & 61.96 & -- \\
OASIS (Qwen2.5-VL-7B) & 67.3 & 52.6 \\
OASIS (Qwen3-VL-8B) & 78.1 & 57.2 \\
Qwen2.5-VL-7B & 60.9 & 50.2 \\
StreamAhead (Qwen2.5-VL-7B) & 69.1 & 51.5 \\
Qwen3-VL-8B & 66.8 & 51.2 \\
StreamAhead (Qwen3-VL-8B) & 76.9 & 53.5 \\
\bottomrule
\end{tabular}%
}
\end{table}

\begin{table}[t]
\centering
\caption{Answer quality on StreamingBench Real-Time
(accuracy, \%). External results are from published
evaluations; StreamAhead uses a randomly sampled subset.}
\label{tab:accuracy-streaming}
\footnotesize
\setlength{\tabcolsep}{5pt}
\renewcommand{\arraystretch}{1.0}
\begin{tabular}{lr}
\toprule
Method & Accuracy (\%) \\
\midrule
VideoLLM-online-8B & 35.99 \\
Dispider & 67.63 \\
StreamAgent-7B & 74.28 \\
Think while Watching & 74.40 \\
SAVEMem & 76.00 \\
HERMES & 79.44 \\
ReKV & 69.10 \\
ReKV + STC & 65.20 \\
OASIS (Qwen2.5-VL-7B) & 70.59 \\
OASIS (Qwen3-VL-8B) & 78.22 \\
StreamAhead (Qwen3-VL-8B) & 79.30 \\
\bottomrule
\end{tabular}
\end{table}

Tables~\ref{tab:accuracy-ovo} and
\ref{tab:accuracy-streaming} evaluate whether independent
segment processing and retrieved representation composition
preserve answer quality. We evaluate StreamAhead on the
full OVO-Bench Perception and Backward subsets and a
randomly sampled subset of approximately half of
StreamingBench Real-Time. Audio-dependent Omni-Source
questions are excluded. External baseline results are
taken from the corresponding publications, including
results compiled in OASIS, rather than reproduced by us.
The Real-Time results therefore provide reference comparisons,
not evaluations on identical queries.

Relative to the corresponding base VLMs, StreamAhead
improves OVO-Bench Perception accuracy from 60.9\% to
69.1\% with Qwen2.5-VL-7B and from 66.8\% to 76.9\%
with Qwen3-VL-8B, while maintaining comparable Backward
accuracy. On StreamingBench Real-Time, StreamAhead
achieves 79.30\%, remaining competitive with published
methods. These results suggest that reusable segment
representations preserve competitive answer quality
without reducing the frames the model sees.

\begin{table}[t]
\centering
\caption{Sensitivity to algorithmic settings.}
\label{tab:sensitivity}
\footnotesize
\setlength{\tabcolsep}{3.4pt}
\renewcommand{\arraystretch}{1.0}
\resizebox{0.8\columnwidth}{!}{%
\begin{tabular}{lrrrr}
\toprule
\multirow{2}{*}{Setting}
& \multicolumn{2}{c}{Accuracy (\%)}
& \multicolumn{2}{c}{Mean TTFT (ms)} \\
\cmidrule(lr){2-3}
\cmidrule(lr){4-5}
& Bwd. & RT & Bwd. & RT \\
\midrule
\multicolumn{5}{l}{\emph{Segmentation}} \\
\textbf{0.5\,s / 15 frames}
& 53.5 & 79.3 & 146.1 & 204.3 \\
0.5\,s / 8 frames
& 52.5 & 78.8 & 123.1 & 154.2 \\
0.5\,s / 4 frames
& 52.0 & 78.5 & 90.1 & 117.9 \\
1.0\,s / 30 frames
& 52.0 & 78.0 & 275.2 & 334.4 \\
1.0\,s / 4 frames
& 49.5 & 77.5 & 89.9 & 98.4 \\
\midrule
\multicolumn{5}{l}{\emph{Retrieval}} \\
Semantic Top-3
& 48.0 & 63.0 & 158.9 & 376.6 \\
Latest + Sem. Top-1
& 51.0 & 80.0 & 120.8 & 164.4 \\
\textbf{Latest + Sem. Top-2}
& 53.5 & 79.3 & 146.1 & 204.3 \\
Latest + Sem. Top-5
& 52.0 & 78.5 & 296.2 & 357.7 \\
\midrule
\multicolumn{5}{l}{\emph{Composition}} \\
\textbf{Concat}
& 53.5 & 79.3 & 146.1 & 204.3 \\
TailRepair
& 54.0 & 79.5 & 266.1 & 344.5 \\
\bottomrule
\end{tabular}%
}
\end{table}

Table~\ref{tab:sensitivity} examines the trade-off between
the frames and segments used, answer accuracy, and query latency.
Processing too few frames can miss useful evidence,
while increasing the number of frames or retrieved
segments does not consistently improve accuracy and
substantially increases TTFT. Latest + Semantic Top-2
achieves the highest Backward accuracy among the
evaluated retrieval policies, whereas Top-1 performs
slightly better on Real-Time. This suggests that many
Real-Time questions can be answered using recent evidence
without extensive historical retrieval. TailRepair
provides modest accuracy gains at considerably higher
latency. We therefore select 0.5-s segments with
15 frames, Latest + Semantic Top-2 retrieval, and
Concat as a balanced default for the
subsequent system evaluation.

\subsection{Video Processing Rate vs. Query Latency}
\label{sec:rate-ttft}

\begin{figure*}[t]
\centering
\begin{minipage}[t]{0.49\textwidth}
    \centering
    \includegraphics[width=\linewidth]
    {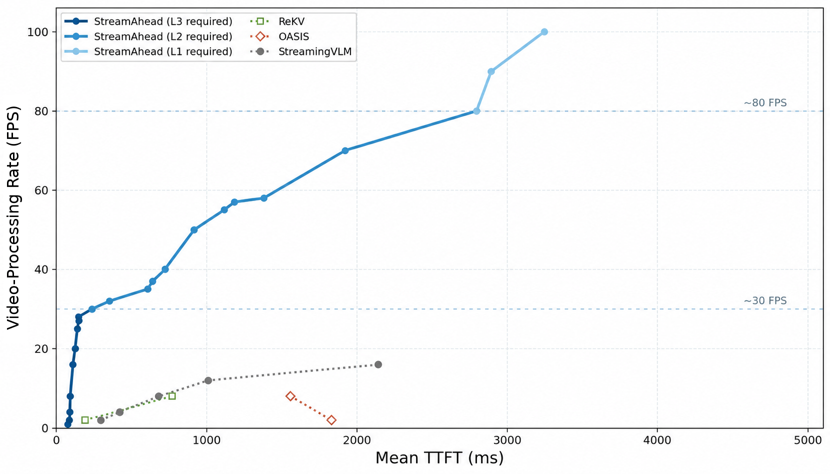}
    \par\small (a) Mean TTFT across video-processing rates.
\end{minipage}
\hfill
\begin{minipage}[t]{0.49\textwidth}
    \centering
    \includegraphics[width=\linewidth]
    {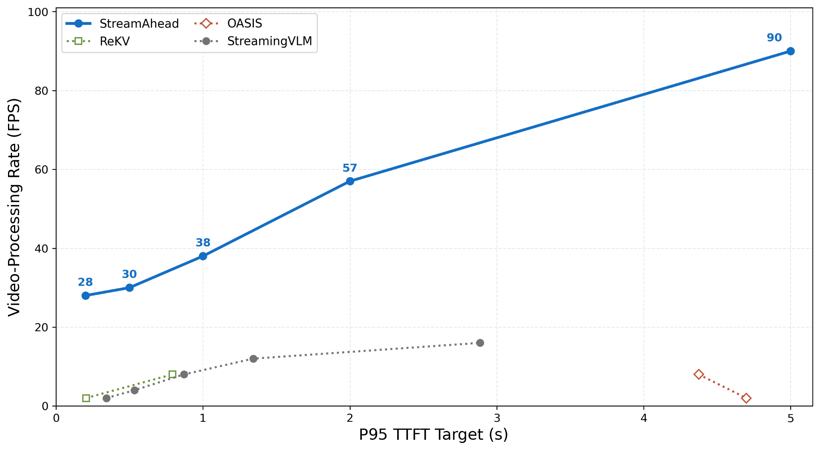}
    \par\small (b) Video-processing rate under P95 TTFT targets.
\end{minipage}
\caption{Video-processing-rate--TTFT trade-off. (a) Mean TTFT across video-processing rates, showing three required representation levels selected by offline profiling: L3 below 30 FPS, L2 from 30 to 80 FPS, and L1 above 80 FPS. (b) Highest video-processing rate satisfying each P95 TTFT target. External methods are shown at their measured TTFT for reference.}
\label{fig:rate-ttft-frontier}
\end{figure*}

We evaluate the rate--latency trade-off through mean TTFT across video-processing rates and the highest rate satisfying each P95 TTFT target. For StreamAhead's rate and TTFT-target sweeps, each replay processes 20\,s of video as 40 segments of 0.5\,s. One query is issued at the end of the trace, and video input stops when the query arrives. Section~\ref{sec:online-scheduling-eval} evaluates queries while video continues to arrive. Figure~\ref{fig:rate-ttft-frontier}(a) shows three operating regions selected by offline profiling. Below 30 FPS, L3 is required, so incoming segments complete their KV representations before queries arrive, enabling low TTFT through direct reuse. From 30 to 80 FPS, L2 is required to maintain the video-processing rate, while additional L3 precomputation uses available capacity. As the input rate increases, fewer segments can reach L3 before query arrival, increasing the remaining query-time computation and TTFT. Above 80 FPS, L1 becomes the required level, further reducing mandatory processing per segment while deferring visual encoding and KV construction until needed. This allows StreamAhead to process up to 100 FPS, although query latency increases under the higher workload. Figure~\ref{fig:rate-ttft-frontier}(b) quantifies this trade-off under P95 TTFT targets: StreamAhead supports 28, 30, 38, 57, and 90 FPS under targets of 0.2, 0.5, 1, 2, and 5\,s, respectively, with P95 processing lag within 1\,s. At 100 FPS, P95 TTFT exceeds the 5-s target, highlighting the need to consider processing rate and query latency jointly. Table~\ref{tab:rate-ttft-comparison} provides a separate online replay comparison. At 8 FPS, StreamAhead achieves 2.5$\times$ and 2.8$\times$ lower mean TTFT than StreamingVLM and ReKV, respectively. At 16 FPS, its TTFT is 5.9$\times$ lower than StreamingVLM, while at 2 FPS it achieves 10.4$\times$ lower TTFT than OASIS. Together, these results show how adaptive precomputation balances continuous video processing against query-time computation across different workload intensities.

\begin{table}[t]
\centering
\caption{Mean TTFT under online replay (ms).}
\label{tab:rate-ttft-comparison}
\footnotesize
\setlength{\tabcolsep}{4pt}
\renewcommand{\arraystretch}{1.05}
\resizebox{0.7\columnwidth}{!}{%
\begin{tabular}{lrrr}
\toprule
Method & 2 FPS & 8 FPS & 16 FPS \\
\midrule
StreamAhead
& \textbf{176.6} & \textbf{273.1} & \textbf{360.8} \\
ReKV
& 191.6 & 771.3 & -- \\
StreamingVLM
& 294.4 & 678.7 & 2140.5 \\
OASIS
& 1830.4 & 1557.2 & -- \\
\bottomrule
\end{tabular}%
}
\end{table}

\subsection{Online Representation Scheduling}
\label{sec:online-scheduling-eval}

\begin{figure}[t]
\centering
\begin{minipage}[t]{0.49\columnwidth}
    \centering
    \includegraphics[width=\linewidth]
    {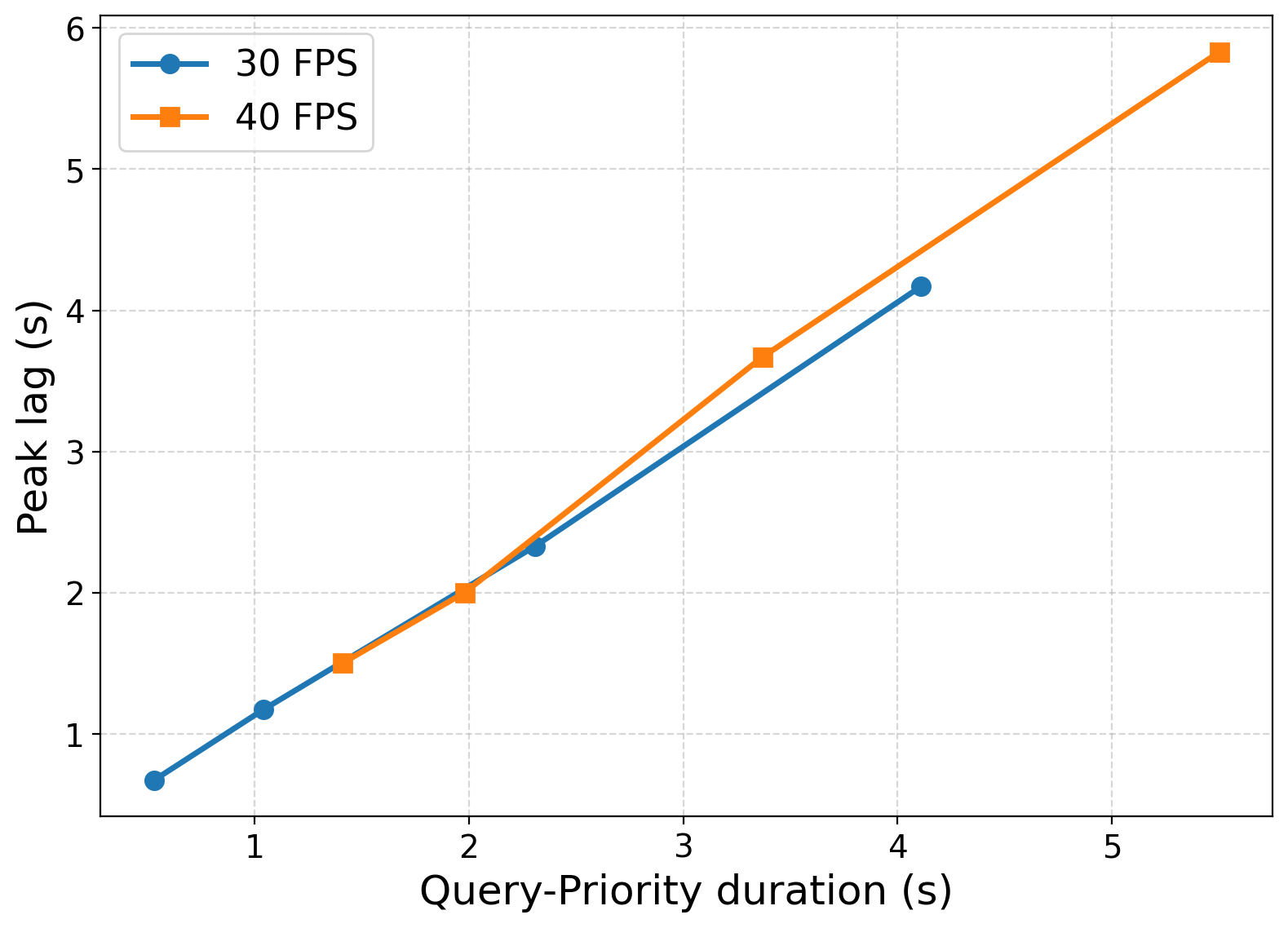}
    \vspace{-0.6em}

    \small (a) Peak processing lag
\end{minipage}%
\hfill
\begin{minipage}[t]{0.49\columnwidth}
    \centering
    \includegraphics[width=\linewidth]
    {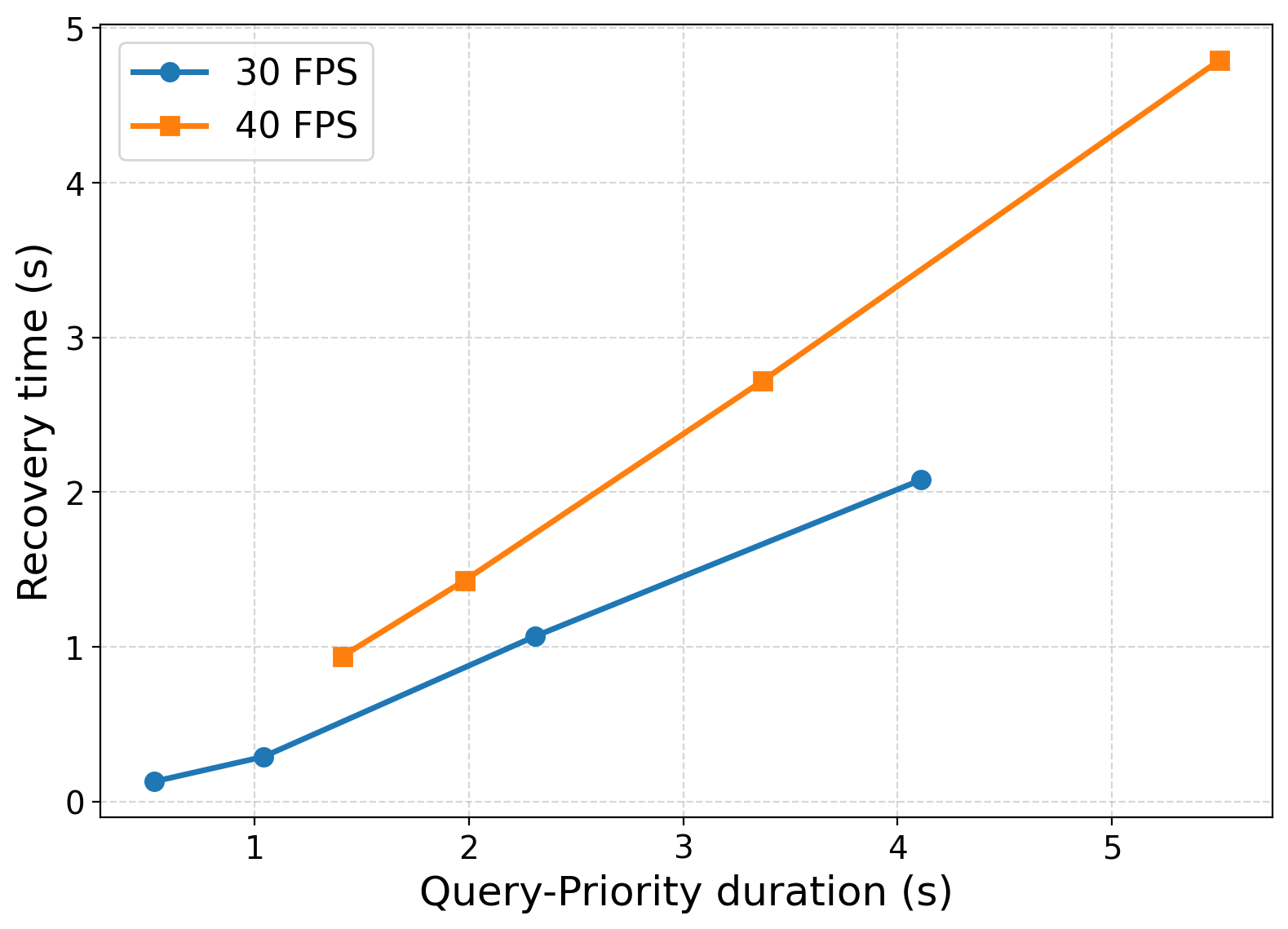}
    \vspace{-0.6em}

    \small (b) Recovery time
\end{minipage}
\caption{Processing lag and recovery for StreamAhead at 30 and
40 FPS under continuous video arrival.
(a) Peak processing lag and (b) recovery time
versus Query-Priority duration.}
\label{fig:online-recovery}
\end{figure}

We evaluate how StreamAhead accumulates and clears temporary
processing lag under continuous video arrival.
We use 0.5-s segments at 30 and 40 FPS, containing 15 and 20 frames,
respectively, with L2 as the required representation level.
Generation budgets of 1, 16, 60, and 128 tokens produce different
Query-Priority durations, while video continues to arrive throughout
query execution and recovery.
During Query-Priority, StreamAhead continues L1 construction for
incoming segments while prioritizing query-critical work on shared
resources, so required L2 processing can temporarily fall behind.
After query execution, Recovery prioritizes pending L2 work and
pipelines it with the processing of newly arriving segments.
Once the required backlog is cleared, the runtime returns to Normal
and resumes additional L3 precomputation.
Figure~\ref{fig:online-recovery} reports StreamAhead's peak processing
lag and recovery time as functions of Query-Priority duration.
At both input rates, longer Query-Priority durations lead to higher
peak lag and longer recovery in the measured runs.
StreamAhead then catches up with the incoming video while new segments
continue to arrive.
These results characterize the temporary lag and subsequent catch-up
behavior of online representation scheduling under continuous input.

\subsection{Runtime Ablation Study}
\label{sec:runtime-ablation}

Table~\ref{tab:runtime-ablation} evaluates three runtime
mechanisms over 180 segments at 30, 40, and 50 FPS.
All other settings, including the 16-GiB GPU cache
and GPU execution optimizations, remain unchanged.
\emph{Required-Only} disables additional L3 precomputation
and maintains only the required L2 level.
\emph{No Query Task Priority} equalizes software task
priorities while retaining CUDA streams.
\emph{No Pipeline} disables CPU--GPU stage overlap.
Required-Only substantially increases TTFT due to
query-time L3 completion, while removing query priority
causes smaller latency increases. More importantly,
No Pipeline meets the 0.5-s P95 processing lag target at
30 FPS but exceeds it at 40 and 50 FPS, reaching
1.0 and 1.5\,s, respectively. In contrast, Full
StreamAhead maintains the lag target across all
rates while achieving the lowest TTFT, demonstrating
the importance of pipeline overlap for sustaining
video processing under higher workloads.

\begin{table}[t]
\centering
\caption{Runtime ablation at 30, 40, and 50 FPS.
TTFT is reported in milliseconds. P95 L2 lag is
reported in seconds. The target lag is 0.5\,s.
All configurations use a 16-GiB GPU cache.}
\label{tab:runtime-ablation}
\resizebox{\columnwidth}{!}{%
\begin{tabular}{lrrrr}
\hline
\textbf{Configuration}
& \textbf{Mean TTFT}
& \textbf{P95 TTFT}
& \textbf{P95 L2 Lag}
& \textbf{L3-ready} \\
\hline
\multicolumn{5}{l}{\textit{30 FPS}} \\
Full StreamAhead         & \textbf{198} & \textbf{330} & 0.5 & 96.2\% \\
Required-Only            & 1181 & 1283 & 0.5 & 0.0\% \\
No Query Task Priority   & 238  & 530  & 0.5 & 94.8\% \\
No Pipeline              & 295  & 453  & 0.5 & 93.7\% \\
\hline
\multicolumn{5}{l}{\textit{40 FPS}} \\
Full StreamAhead         & \textbf{622} & \textbf{830} & 0.5 & 62.7\% \\
Required-Only            & 1586 & 1732 & 0.5 & 0.0\% \\
No Query Task Priority   & 682  & 851  & 0.5 & 61.3\% \\
No Pipeline              & 859  & 1068 & \textbf{1.0} & 58.2\% \\
\hline
\multicolumn{5}{l}{\textit{50 FPS}} \\
Full StreamAhead         & \textbf{1011} & \textbf{1769} & 0.5 & 33.5\% \\
Required-Only            & 2201 & 2393 & 0.5 & 0.0\% \\
No Query Task Priority   & 1074 & 1810 & 0.5 & 32.2\% \\
No Pipeline              & 1216 & 1952 & \textbf{1.5} & 31.5\% \\
\hline
\end{tabular}%
}
\end{table}





\subsection{Representation Reuse and Memory Placement}
\label{sec:memory-eval}

We evaluate the memory hierarchy by separating \emph{representation level}, which determines remaining query computation, from \emph{placement}, which determines whether completed states must be transferred. L3 representations require 33.6\,MiB per segment with one frame and 841.3\,MiB with 30 frames, motivating a bounded 16-GiB GPU cache backed by CPU storage. Under Latest + Semantic Top-2 retrieval, approximately 35\% of retrieved historical L3 states are served directly from GPU memory. Table~\ref{tab:representation-state} holds the retrieved segments fixed while varying their states: when both are GPU-resident L3, TTFT is 90.1\,ms, compared with 254.5\,ms when neither historical L3 is in GPU memory; when one is only at L2, query-time completion increases TTFT to 490.1\,ms. These results show why reusable computation and its CPU--GPU placement must be managed separately under a fixed GPU memory budget.

\begin{table}[t]
\centering
\caption{TTFT under different historical representation states (ms).}
\label{tab:representation-state}
\footnotesize
\setlength{\tabcolsep}{3pt}
\renewcommand{\arraystretch}{1.0}
\resizebox{0.9\columnwidth}{!}{%
\begin{tabular}{lrrrr}
\toprule
Historical state & Evict D2H & L3 H2D & L2$\to$L3 & TTFT \\
\midrule
0/2 L3 on GPU & 86.1 & 59.9 & 0.0 & 254.5 \\
1/2 L3 on GPU & 39.8 & 29.9 & 0.0 & 176.1 \\
2/2 L3 on GPU & 0.0 & 0.0 & 0.0 & 90.1 \\
1 CPU L2 + 1 GPU L3 & 36.6 & 0.0 & 339.3 & 490.1 \\
\bottomrule
\end{tabular}%
}
\end{table}

\section{Conclusion}
\label{sec:conclusion}

This paper presented StreamAhead, an algorithm--system co-design for real-time
streaming video QA. StreamAhead processes video segments independently at three
representation levels and decides at runtime how much computation to complete
for each segment before a query arrives. Its profile-guided scheduler and
concurrent CPU--GPU runtime keep pace with the incoming video while completing
KV caches only for retrieved segments. StreamAhead maintains competitive
accuracy with up to 10.4$\times$ lower TTFT than state-of-the-art methods, and
sustains 57\,FPS with P95 TTFT under 2\,s, setting a new Pareto frontier between
video-processing rate and TTFT. These results show that how much computation to
complete before a query arrives should be a runtime decision.
StreamAhead has limitations. Because segments are encoded independently, they
do not attend to one another, and suffix recomputation adds only limited
cross-segment interaction. Our system evaluation uses a single GPU, and rates
above the source frame rate repeat source frames. Offline profiles also assume a
stable workload; extending StreamAhead to multiple GPUs and changing workloads
is future work.

\section*{Acknowledgment}
AI assistants (ChatGPT and Claude) were used to draft and edit text, write
code, and manage experiments. The authors take full responsibility for all content.

\bibliographystyle{IEEEtran}
\bibliography{references}

\end{document}